\documentclass[twocolumn]{aastex631}

\usepackage{amsmath}
\usepackage{CJKutf8}
\received{xxx}
\revised{xxx}
\accepted{xxx}

\begin{document}
\begin{CJK*}{UTF8}{gbsn}

\title{Forward and Reverse Shock Emission from Relativistic Jets with Arbitrary Angular and Stratified Radial Profiles}

\correspondingauthor{Hao Wang}
\email{haowang@pmo.ac.cn; haowang.astro@gmail.com}

\author[0000-0002-0556-1857]{Hao Wang (王灏)}
\affiliation{Key Laboratory of Dark Matter and Space Astronomy, Purple Mountain Observatory, \\
Chinese Academy of Sciences, Nanjing 210023, China}

\author[0000-0003-2915-7434]{Hao Zhou}
\affiliation{Key Laboratory of Dark Matter and Space Astronomy, Purple Mountain Observatory, \\
Chinese Academy of Sciences, Nanjing 210023, China}

\author[0000-0002-8966-6911]{Yi-Zhong Fan}
\affiliation{Key Laboratory of Dark Matter and Space Astronomy, Purple Mountain Observatory, \\
Chinese Academy of Sciences, Nanjing 210023, China}
\affiliation{School of Astronomy and Space Science, University of Science and Technology of China, \\
Hefei 230026, China}

\author[0000-0002-9758-5476]{Da-Ming Wei}
\affiliation{Key Laboratory of Dark Matter and Space Astronomy, Purple Mountain Observatory, \\
Chinese Academy of Sciences, Nanjing 210023, China}
\affiliation{School of Astronomy and Space Science, University of Science and Technology of China, \\
Hefei 230026, China}



\begin{abstract}
Gamma-ray bursts are expected to be generated by structured jets, whose profiles significantly impact their afterglow emission. Previously, we developed a numerical code \texttt{jetsimpy}, to model the afterglow of jets with arbitrary angular profiles. In this study, we extend the code to incorporate a stratified radial profile, enabling it to model jets with arbitrary axisymmetric two-dimensional structures. The radial profile leads to the formation of a reverse shock. We modeled the shock system using an energy conservation prescription, which differs from the pressure balance approach. This leads to remarkably different predictions for reverse shock emission. In particular, we find that the reverse shock emission in the thin shell case is significantly overestimated in analytic models. We also explore the off-axis reverse shock emission from structured jets, where the cores belong to thick shell cases and the wings belong to thin shell cases. We have confirmed the prediction that off-axis observers may see a thin-to-thick transition, but we find that the light curve morphology is hard to distinguish from pure thin or thick shell cases. A radial profile also introduces hydrodynamic energy injection. As such, our code can naturally apply to refreshed shock cases, where the modeling of kilonova afterglows is demonstrated as an example. To validate our method, we fit the optical flash of GRB 990123, showing good agreement with the data. The upgraded \texttt{jetsimpy} provides unprecedented flexibility in modeling the afterglow emission of jets with various profiles, including those derived from general relativistic magnetohydrodynamic simulations.

\end{abstract}



\section{Introduction} \label{sec:intro}
Gamma-ray bursts (GRBs) are the most powerful explosions in the universe, produced by narrow ultra-relativistic jets. 
After the prompt gamma-ray emission phase, the jet interacts with the ambient medium, giving rise to the afterglow emission. This interaction develops a forward-reverse shock system, in which the forward shock propagates outward through the surrounding matter, and the reverse shock travels backward through the ejecta. These shocks accelerate electrons and generate synchrotron radiation. The reverse shock emission usually peaks very rapidly. Under the assumption of certain parameters, it is thought to be responsible for the optical flash during the very early afterglow phase. The forward shock emission can extend from the radio to the gamma-ray bands and may last for years.

Both the dynamical evolution of a GRB jet and its afterglow emission depend on the jet's profile. For an arbitrary jet configuration under the assumption of axial symmetry, we can decompose it into an angular profile (e.g., \citealt{1998ApJ...499..301M,2002MNRAS.332..945R,2003ApJ...591.1086G,2003ApJ...591.1075K}) and a stratified radial profile (e.g., \citealt{1998ApJ...503..314P,1998ApJ...496L...1R,2000ApJ...535L..33S}). The angular profile determines how the emission varies in different directions, and the radial profile determines how energy is gradually injected into the blast region. 

The jet angular profile usually consists of a ``core" region where the energy distribution is approximately uniform and a ``wing" region where the energy decays at larger latitudes. The angular profile of this ``wing" region becomes important when the observer’s line of sight is misaligned from the jet core (though a very flat wing can even affect on-axis observers, see \citealt{2023MNRAS.524L..78G}).
This is exactly the case for the GRB detected in the neutron star merger event GW170817 (\citealt{2017PhRvL.119p1101A,2017ApJ...848L..12A,2017ApJ...848L..13A}; see \citealt{2021ARA&A..59..155M} for a review), where the observing angle is believed to be much larger than the jet’s half-opening angle \citep{2018ApJ...857..128J,2017ApJ...848L..25H, 2017ApJ...848L..20M, 2018ApJ...856L..18M, 2017Natur.551...71T, 2018MNRAS.478L..18T, 2020MNRAS.498.5643T, 2018ApJ...863L..18A, 2018MNRAS.481.2711G, 2018PhRvL.120x1103L, 2018Natur.561..355M, 2019ApJ...883L...1F, 2019ApJ...886L..17H, 2019ApJ...870L..15L, 2019ApJ...880L..23W}. Unlike typical GRB afterglows, which are mostly on-axis, this off-axis afterglow exhibits a rising light curve. This occurs because the beaming effect weakens as the jet decelerates, and the bright core gradually becomes visible. The features of off-axis afterglow depend sensitively on the jet angular profile and observing angle (see \citealt{2020MNRAS.493.3521B} for an analytic modeling). On the other hand, when the angular profile is very shallow, the angular profile becomes important even for an on-axis observers \citep{2022MNRAS.515..555B,2023MNRAS.524L..78G}. The major challenge in modeling a jet with such a structure lies in its spreading phase \citep{1999ApJ...525..737R, 1999ApJ...519L..17S,1999ApJ...525..737R,2012MNRAS.421..570G, 2020ApJ...896..166R, 2023MNRAS.520.2727N}, where the jet expands laterally due to its internal pressure. The lateral interaction between fluid elements with different energies involves nonlinear hydrodynamic processes, which are difficult to describe by analytic methods. The most accurate way to model this phase is through hydrodynamic simulation (e.g., \citealt{2003ApJ...591.1075K,2009ApJ...698.1261Z,2010ApJ...722..235V,2012ApJ...749...44V,2018ApJ...865...94D,2024MNRAS.531.1704G}). Recently, we have developed a new method based on a simplified simulation approach \citep{2024ApJS..273...17W}. This method allows for the simulation of a jet with an arbitrary angular profile with accuracy similar to full simulation while maintaining the computational efficiency of semi-analytic methods. Based on this method, we have developed a numerical tool \texttt{jetsimpy}\footnote{\url{https://github.com/haowang-astro/jetsimpy}} which has been provided to the community.

The jet radial profile, on the other hand, has two effects. First, the finite ejecta shell thickness allows the reverse shock to travel through the shell until it reaches the ejecta tail. This reverse shock generates an emission component, which in some cases exceeds the forward shock emission at early times. This emission is widely believed to be the origin of early optical flashes observed in some GRB afterglows \citep{1999ApJ...513..669K,2000ApJ...542..819K,2000ApJ...545..807K,2002ChJAA...2..449F,2003ApJ...595..950Z,2003ApJ...597..455K,2004A&A...424..477F,2005ApJ...628..315Z,2005ApJ...628..867F}. Second, if the velocity distribution of the ejecta is significantly extended, the expansion of the ejecta becomes homogeneous, and the reverse shock is long-lasting. As the blastwave decelerates, the slow components that lagged behind will continuously refresh the shocked regions and increase the energy of the blastwave. This effect may account for the early shallow decay phase (e.g., \citealt{2006ApJ...642..354Z,2007ApJ...665L..93U,2016MNRAS.457L.108B}) or the late time rebrightening observed in some afterglows (e.g., \citealt{2003Natur.426..138G,2015ApJ...814....1L,2018ApJ...862...94L,2019ApJ...883...48L,2023MNRAS.525.5224M}). It is also used to model the afterglow emission of the blastwave generated by kilonova ejecta surrounding a GRB \citep{2011Natur.478...82N,2013MNRAS.430.2121P,2017ApJ...848L..21A,2018ApJ...867...95H,2018ApJ...869..130R}.
 
A misaligned GRB jet with a radial profile is less well studied when combined with its angular profile.  This is mainly because significantly misaligned prompt emission will be less bright at cosmological distances (e.g., \citealt{2019MNRAS.482.5430B}) and will not trigger follow-up observations. However, the afterglow emission in this situation could still be detectable, which has long been proposed as ``orphan afterglows” (e.g., \citealt{1997ApJ...487L...1R,2002ApJ...579..699N}). Recent studies have found that off-axis reverse shock emission may still exceed forward shock emission in the early afterglow phase \citep{2024MNRAS.528.2066P,2024ApJ...977..123P,2025arXiv250212757A}. In certain situations, the combination of bright reverse shock emission and forward shock emission will exhibit a double peak feature \citep{2025arXiv250212757A}, which has not yet been detected. In recent years, optical sky surveys have already identified a few orphan afterglow candidates (e.g., \citealt{2025MNRAS.538..351S,2025MNRAS.537.2362P,2023MNRAS.523.4029L}). The operation of the Einstein Probe \citep{2022hxga.book...86Y} may also help identify off-axis prompt emission of GRBs. Together, these telescopes shed light on the discovery of off-axis reverse shock emission.

In this paper, we extend our previous work by further developing \texttt{jetsimpy} to model GRB afterglows for jets with stratified radial profiles. This upgrade significantly enhances the code, which enables it to model the emission from both forward and reverse shocks produced by GRB jets with arbitrary axisymmetric two-dimensional configurations. Additionally, it can naturally model the kilonova afterglows within the framework of refreshed shocks. The flexibility of the code enables it to incorporate jet profiles derived from general relativistic magnetohydrodynamic (GRMHD) simulations as initial conditions and study their afterglow emission. As such, the upgraded \texttt{jetsimpy} version will serve as a valuable tool for the relevant community.

This paper is organized as follows. In section \S\ref{sec:methods} we introduce our method to include the radial profiles. In section \S\ref{sec:cases} we apply our model in some typical cases, and compare them to previous studies. In section \S\ref{sec:grb990123} we apply our model to fit the reverse shock emission of GRB 990123 and validate our approach. We summarize and discuss our results in \S\ref{sec:summary}.

\section{Methods} \label{sec:methods}

\subsection{The forward-reverse shock system} \label{subsec:shock system}
\begin{figure}[ht!]
\centering
\includegraphics[width=\columnwidth]{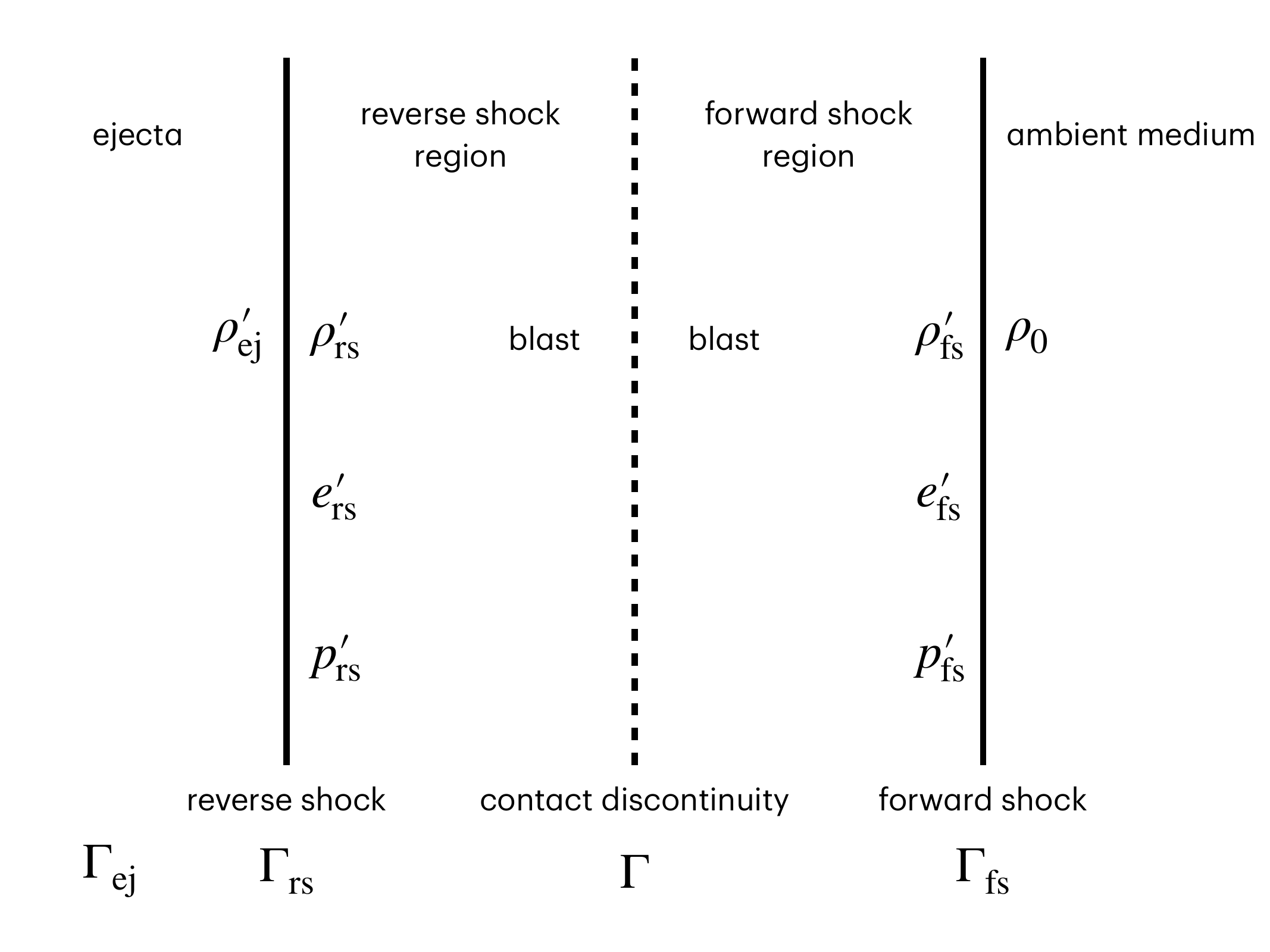}
\caption{An illustration of the forward-reverse shock system and the labeling of the relevant thermodynamic quantities.}
\label{fig:shock}
\end{figure}

The interaction between a relativistic jet and its surrounding medium will develop a forward-reverse shock system consisting of three discontinuities. The interface between the two fluids forms a contact discontinuity, the interaction propagating in the ambient matter forms a forward shock, and the interaction propagating in the ejecta forms a reverse shock. The shocks and the contact discontinuity separate the system into four regions: the ambient medium, the shocked ambient medium, the shocked ejecta, and the ejecta. Throughout this paper, we will refer to the region of shocked ambient medium as the forward shock region and the region of shocked ejecta as the reverse shock region. The thermodynamic properties of each region can be described by its density $\rho'$, rest mass excluded energy density $e'$, and pressure $p'$, where each of the values is measured in the corresponding comoving frame. The dimensionless velocity and Lorentz factor of each region are labeled as $\beta$ and $\Gamma$, which are measured in the burster frame. In this work, we assume the ambient medium is cold and at rest. We also assume that the internal energy of the ejecta has mostly converted to kinetic energy after the prompt emission phase, indicating that the ejecta is also cold (see \citealt{2017ApJ...846...54P} for the case of hot ejecta). We assume the ejecta is not magnetized and leave the study of magnetized ejecta for future work. These assumptions imply that the pressures of the ambient medium and the ejecta are negligible. As such, the only hot regions are the forward and reverse shock regions, which is referred to as the blast region as a whole. We assume the velocity in the blast region is uniform. The forward-reverse shock system is illustrated in Figure \ref{fig:shock}, where we have labeled all relevant values in each region. We notify readers that the labeling in this work is different from our previous work \citep{2024ApJS..273...17W}, where the properties of the shocked ambient medium were labeled as ``sw.” Here, we label them as ``fs” for better clarity with respect to the label ``rs” in the reverse shock region.

The boundary condition of the blast region is determined by the shock jump conditions. Here, we follow \citealt{2011ApJ...733...86U}, who derived the jump conditions assuming a trans-relativistic equation of state (\citealt{2007MNRAS.378.1118M}; see also \citealt{1977ApJ...218..377W,1995ApJ...455L.143S,2017ApJ...846...54P}). At the forward shock, the jump conditions are:
\begin{align}
    & \rho'_{\rm fs} = 4\Gamma \rho_0, \\
    & e'_{\rm fs} = 4\Gamma(\Gamma-1)\rho_0 c^2, \\
    & p'_{\rm fs} = \frac{4}{3}(\Gamma^2-1) \rho_0 c^2.
\end{align}
At the reverse shock, the jump conditions are
\begin{align}
    & \rho'_{\rm rs} = 4\Gamma_{34} \rho'_{ej}, \\
    & e'_{\rm rs} = 4\Gamma_{34}(\Gamma_{34}-1)\rho'_{\rm ej} c^2, \\
    & p'_{\rm rs} = \frac{4}{3}(\Gamma_{34}^2-1) \rho'_{\rm ej} c^2,
\end{align}
where $\Gamma_{34}=\Gamma\Gamma_{\rm ej}(1-\beta\beta_{\rm ej})$ is the relative Lorentz factor between the ejecta and the reverse shock region\footnote{In some literature, this value is defined as $\Gamma_{34}=(\Gamma/\Gamma_{\rm ej}+\Gamma_{\rm ej}/\Gamma)/2$, which is the ultra-relativistic limit of our expression.}. In this system there are three radius of interest: the radius of the forward shock $R_{\rm fs}$, the radius of the contact discontinuity $R_{\rm cd}$, and the radius of the reverse shock $R_{\rm rs}$. Their velocities can also be determined by the jump conditions, which are
\begin{align}
    & \beta_{\rm fs} = \frac{4\beta\Gamma^2}{4\Gamma^2-1} = \beta + \frac{\beta}{\Gamma^2(3+\beta^2)}, \\
    & \beta_{\rm cd} = \beta, \\
    & \beta_{\rm rs} = \beta - \frac{\beta_{34}}{\Gamma^2(3 - \beta\beta_{34})}.
\end{align}

To solve for the six thermodynamic properties, as well as the Lorentz factor of the blast region, one needs seven equations. However, the jump conditions only provide six of them. An additional condition is required to uniquely determine these values. A common assumption is the pressure balance at the contact discontinuity (e.g., \citealt{1995ApJ...455L.143S}). If the thermodynamic properties in the forward and reverse shock regions are assumed to be uniform, this condition implies that $p'_{\rm fs}=p'_{\rm rs}$. However, many studies have shown that this assumption breaks down the energy conservation of the blast (e.g., \citealt{2006ApJ...651L...1B,2011ApJ...733...86U}). In fact, the pressure continuously varies in the blast. Even the pressure is balanced at the contact discontinuity, it is not constant throughout the blast region. If, to first-order accuracy, a uniform density distribution is assumed in the forward and reverse shock regions, respectively, the pressure balance assumption needs to give way to energy conservation. Energy conservation is naturally satisfied in our approach because the evolution equations are derived from the Euler equations, where energy is strictly conserved.

\subsection{The evolution equations} \label{subsec:dynamics}
The derivation of the evolution equations closely follows our previous work \citep{2024ApJS..273...17W}, to which the readers may refer for more details. We start from the Euler equations, which are simply conservation laws for energy and mass. These equations are:
\begin{align}
    & T^{\mu\nu}_{;\mu}=0, \\
    & J^{\mu}_{;\mu} = 0,
\end{align}
where $T^{\mu\nu}$ is the energy momentum tensor, and $J^{\mu}$ is the mass current. For a forward-reverse shock system, we would like to conserve the total energy of the blast region without bothering with the interaction at the contact discontinuity. We also would like to conserve the mass in the forward and reverse shock regions, respectively. As such, we can construct the total energy momentum tensor $T^{\mu\nu}$, the mass current of the forward shock region $J_{\rm fs}^{\mu}$, and the mass current of the reverse shock region $J_{\rm rs}^{\mu\nu}$ as follows:
\begin{align}
    & T^{\mu\nu} = h'_bu^{\mu}u^{\nu} + g^{\mu\nu}(p'_{\rm fs}+p'_{\rm rs}), \\
    & J_{\rm fs}^{\mu} = \rho'_{\rm fs}u^{\mu}, \\
    & J_{\rm rs}^{\mu} = \rho'_{\rm rs}u^{\mu},
\end{align}
where $h'_b$ is the enthalpy of the blast defined by
\begin{equation}
    h'_b = e'_{\rm fs} + p'_{\rm fs} + \rho'_{\rm fs}c^2 + e'_{\rm rs} + p'_{\rm rs} + \rho'_{\rm rs}c^2.
\end{equation}
The metric tensor is defined in the spherical coordinate. The four-velocity $u^{\mu}$ is defined as
\begin{equation}
    u^{\mu} = (\Gamma, \beta_{\rm r}\Gamma, \frac{\beta_{\theta}\Gamma}{r}, 0).
\end{equation}
The above definition of the energy momentum tensor is the sum of the two shock regions. This helps to avoid details of the interaction between the two regions and to focus on total energy conservation. This sum can be regarded as a smooth interpolation of the two extreme cases: the forward shock dominated case and the reverse shock dominated case. 

In our previous work, the terms in $T^{\mu\nu}$ and $J^\mu$ were multiplied by a constant coefficient to calibrate their values with exact self-similar solutions. However, when a reverse shock is present, this calibration makes no sense, because the blast region is no longer self-similar. In this work, we neglect the calibration process, and the coefficient is assumed to be 1. We emphasize that the calibration coefficient only introduces an order of unity correction to the dynamics of the system, and it has no impact on the scaling behavior. This calibration can still be retained if the radial profile is absent.

Following our previous work, we approximate the blast region as an infinitely thin two-dimensional surface. This approximation is valid even if the ejecta shell is thick, because our simulation is performed only in the region between the forward and reverse shocks, which is still thin in the ultra-relativistic limit. In the thin shell limit, the thermodynamic properties can be approximated by Dirac delta functions. The Euler equations can then be analytically integrated from the reverse shock to the forward shock along the radial dimension, which simplifies them to a one-dimensional equation. The details of the delta function approximation and the radial integral are described in \citep{2024ApJS..273...17W}. After the radial integral, the Euler equations become the following form:
\begin{equation}\label{eq:evolution_equation}
    \frac{\partial {\bf U}}{\partial t} + \frac{1}{\sin\theta}\frac{\partial ({\bf F}\sin\theta)}{\partial \theta} + {\bf S_{\rm fs}} + {\bf S_{\rm rs}} = 0.
\end{equation}
The conserved variables ${\bf U}$, the flux vector ${\bf F}$, and the source terms $\bf S_{\rm fs}$ and $\bf S_{\rm rs}$ are defined as
\begin{equation}
    {\bf U} = 
    \begin{pmatrix}
        E_b \\
        \beta_{\theta}H_b \\
        M_{\rm fs} \\
        M_{\rm rs} 
    \end{pmatrix},
\end{equation}
\begin{equation}
    {\bf F} = \frac{c}{R_{\rm fs}}
    \begin{pmatrix}
        \beta_{\theta}H_b \\
        \beta_{\theta}^2 H_b + P_b \\
        \beta_{\theta} M_{\rm fs} \\
        \beta_{\theta} M_{\rm rs}
    \end{pmatrix},
\end{equation}
\begin{equation}
    {\bf S_{\rm fs}} = 
    \begin{pmatrix}
        - \frac{\partial R_{\rm fs}}{\partial t}\rho_0 R_{\rm fs}^2 \\
        (\beta_{\theta}\beta_{\rm r}H_b - \frac{\cos\theta}{\sin\theta}P_b)\frac{c}{R_{\rm fs}} \\
        - \frac{\partial R_{\rm fs}}{\partial t}\rho_0 R_{\rm fs}^2 \\
        0
    \end{pmatrix}, \\
\end{equation}
\begin{equation}
    {\bf S_{\rm rs}} = 
    \begin{pmatrix}
        - (\beta_{\rm ej}-\beta_{\rm rs})\Gamma_{\rm ej}^2\rho'_{\rm ej}R_{\rm rs}^2c \\
        0 \\
        0 \\
        - (\beta_{\rm ej}-\beta_{\rm rs})\Gamma_{\rm ej}\rho'_{\rm ej}R_{\rm rs}^2c
    \end{pmatrix}.
\end{equation}
In the above equations, $E_b$, $H_b$, $P_b$ are the radially integrated energy, enthalpy, and pressure of the blast region. $M_{\rm fs}$, and $M_{\rm rs}$ are the radially integrated mass in the forward and reverse shock regions, respectively. The energy, enthalpy, and pressure are divided by $c^2$ to align their unit with mass. The shock jump conditions lead to the following relations between these values
\begin{align}
    & E_b = \Gamma^2(1+\frac{1}{3}\beta^4)M_{\rm fs} + \Gamma\Gamma_{34}(1+\frac{1}{3}\beta^2\beta_{34}^2)M_{\rm rs}, \label{eq:energy} \\
    & P_b = \frac{1}{3}\beta^2 M_{\rm fs} + \frac{1}{3}\beta_{34}^2\frac{\Gamma_{34}}{\Gamma}M_{\rm rs}, \\
    & H_b = E_b + P_b.
\end{align}
We can easily see that the evolution equations are similar to those in our previous work. Differences in the definitions of energy and pressure arise from the contributions of the hot reverse shock region. The difference in the source term (i.e., the additional $\bf S_{\rm rs}$ term) originates from the accumulation of ejecta matter by the reverse shock. These equations reduce to the original form presented in our previous work when the reverse shock disappears (i.e., $\Gamma{34}=1$ and $\beta_{34}=0$). When the lateral expansion effect is ignored, our formalism is similar to the ``mechanic model" first explored in \citet{2006ApJ...651L...1B} and other similar works (e.g., \citealt{2013MNRAS.433.2107N,2022MNRAS.513.4887Z,2023MNRAS.524L..78G}; see also \citealt{2021MNRAS.507.1788A} for magnetized outflows).

Some multidimensional numerical studies have found that the contact discontinuity could be Rayleigh-Taylor unstable (e.g., \citealt{2013ApJ...775...87D}), which can completely disrupt the interface, causing the thermodynamic variables to vary continuously in the blast region. These hydrodynamical instabilities are beyond the scope of our model. However, the radially integrated variables in our method can still be regarded as Riemann sum approximations to the actual results, because the jump conditions are not affected by the instabilities.

In these equations, the lateral expansion is mechanically driven by the pressure term $P_b$, which contains the pressure in both the forward and reverse shock regions. However, the pressure difference between the two regions could induce different expansion speeds, which conflicts with our assumption of a single value of $\beta_{\theta}$ for the blast region. This velocity difference could even trigger hydrodynamic instabilities, which are beyond the scope of this work. 
For the specific case of GRB afterglows, lateral expansion becomes important only when the outflow has decelerated to mildly relativistic speeds, due to causality considerations. Therefore, it typically occurs after the deceleration timescale, when the pressure in the forward shock region starts to dominate the overall pressure in the blast region. This consideration is consistent with the findings in multidimensional numerical studies (e.g., \citealt{2013ApJ...775...87D}). For this reason, in this study we assume that lateral expansion is powered purely by pressure in the forward shock region.
We neglect the contribution of the reverse shock to the total pressure and only consider its contribution to the total energy. As such, the $P_b$ term is manually adjusted to
\begin{equation}
    P_b \equiv \frac{1}{3}\beta^2 M_{\rm fs}. 
\end{equation}
We find that this adjustment better stabilizes the hydrodynamic simulation. For all cases investigated in this work, we find the synthetic light curves show no practical differences compared to those generated using the original expression of $P_b$. We also notice that theoretically, when the outflow is initially very slow, lateral spreading may occur while the reverse shock region still dominates the blastwave. In this situation, the pressure in the reverse shock region may not be negligible. To increase flexibility, the original $P_b$ expression is remained as a viable option to solve the blastwave dynamics in our code. However, we caution that the validity of our modeling in this specific regime is not guaranteed and full hydrodynamic simulations should be preferred.

To close the equations, we also need the evolution of the radius of discontinuities:
\begin{align}
    & \frac{\partial R_{\rm fs}}{\partial t} = \beta_{\rm fs}c - \frac{\partial R_{\rm fs}}{\partial \theta}\frac{\beta_{\theta}c}{R_{\rm fs}}, \\
    & \frac{\partial R_{\rm cd}}{\partial t} = \beta_{\rm cd}c - \frac{\partial R_{\rm fs}}{\partial \theta}\frac{\beta_{\theta}c}{R_{\rm fs}}, \\
    & \frac{\partial R_{\rm rs}}{\partial t} = \beta_{\rm rs}c - \frac{\partial R_{\rm fs}}{\partial \theta}\frac{\beta_{\theta}c}{R_{\rm fs}},
\end{align}
where we have assumed that the angular velocity of the reverse shock and contact discontinuity align with the forward shock. 

Now, the blast dynamics can be numerically solved once the boundary conditions (i.e., $\rho_0$, $\rho_{\rm ej}$, and $\beta_{\rm ej}$) are specified. The numerical scheme is similar to that in our previous work, except that the additional reverse shock terms introduce modifications to the eigenvalues of the equation system. We summarize the modified eigenvalues in Appendix \ref{appendix:numerical_scheme}. The detailed numerical scheme can be found in \citet{2024ApJS..273...17W} (see Appendix A therein).

\subsection{The cooling of reverse shock region}
During the evolution of the blastwave, a special situation occurs when the reverse shock passes through the end of the ejecta shell. This time is generally referred to as the crossing timescale, after which the reverse shock disappears and a rarefaction wave develops. If we abruptly remove the contribution of the reverse shock, two difficulties will arise in our model. The first difficulty is that a sudden acceleration of the blastwave occurs at the crossing timescale. The abrupt removal of the reverse shock leads to a sudden cooling of the region. As a result, the energy conservation forces the region to convert all its internal energy into kinetic energy, showing as a sudden acceleration. The second difficulty is that the width of the reverse shock region becomes infinite after the crossing time. This happens because the mass in the region is finite, but the density predicted by the jump conditions drops to zero. These difficulties originate from the infinitely thin shell approximation for the blast region in our method, where the radial dimension is unresolved. An accurate description of the transition from reverse shock to rarefaction wave and its subsequent evolution is beyond the scope of this work. In this section, we introduce phenomenological workarounds to resolve these issues.

To address the first difficulty, it is necessary to continuously evolve $\Gamma_{34}$ before and after the crossing timescale. In practice, the reverse shock region gradually cools and becomes subdominant within the blastwave. However, the cooling of the reverse shock region is a non-linear process that can only be accurately modeled through full hydrodynamic simulations, which is beyond the scope of this work. This challenge appears in all analytic and semi-analytic models and is not unique to our approach. Previous studies have proposed a workaround by introducing a free parameter $g$ and assuming $\Gamma \propto r^{-g}$ during this phase (e.g., \citealt{1999MNRAS.306L..39M,2000ApJ...542..819K}). However, this method is difficult to implement in our approach because it doesn't ensure the energy conservation.

Motivated by previous works, we propose a new workaround. We observe that the relative Lorentz factor $\Gamma_{34}$ reflects the temperature of the reverse shock region and depends only on $\Gamma_{\rm ej}$ and $\Gamma$. Therefore, we assume that the ejecta segment is followed by an imagined tail with zero density but non-zero velocity. The finite $\Gamma_{\rm ej}$ allows for a gradual decrease for $\Gamma_{34}$, while the zero ejecta density ensures conservation of both energy and mass. As a result, after the reverse shock crossing, there is still an ``imagined" reverse shock traveling in this tail. We further assume that the velocity of the tail decays as the imagined reverse shock lags behind the ejecta shell. Specifically, we assume
\begin{equation}
    u_{\rm ej}=u_{\rm min}\left(\frac{\beta_{\rm max} ct-R_{\rm rs}}{\beta_{\rm max}ct-\beta_{\rm min}ct}\right)^{-g}, 
\end{equation}
where $u_{\rm min}$ is the four-velocity of the ejecta tail, $\beta_{\rm max}$ and $\beta_{\rm min}$ are the dimensionless velocities of the ejecta head and tail, respectively, and $g$ is a phenomenological cooling coefficient. By adjusting $g$, this workaround controls how quickly the reverse shock region cools without violating conservation laws. Our tests show that when $g>1$, the cooling process is rapid, while when $g<1$, the cooling process is slow, and the reverse shock emission becomes long-lived. Therefore, the value of $g$ of interest should be around 1.

After the reverse shock crossing, the density estimated by the jump conditions becomes zero, leading to an infinite estimated width for the reverse shock region. To realize this second difficulty, we first revisit the approach that we previously adopted to estimate the blast shell width. The width of the forward shock region is estimated by comparing its density and mass: $\Delta'_{\rm fs}= M_{\rm fs}/R_{\rm fs}^2\rho'_{\rm fs}$, where the width is measured in the comoving frame. The reverse shock region can be treated similarly: $\Delta'_{\rm rs}\approx M_{\rm rs}/R_{\rm rs}^2\rho'_{\rm rs}$. After the reverse shock crossing, the absence of ejecta causes $\rho'_{\rm ej}$ to drop to zero, making this estimation invalid.

To find a workaround, we note that in our framework there exists an alternative method to estimate the shell width, which is to measure the distance between the contact discontinuity and the reverse shock: $\Delta'_{\rm rs}=\Gamma(R_{\rm cd}-R_{\rm rs})$. This estimation remains valid after the crossing time as long as the above $g$ workaround is adopted, because the imagined reverse shock continuously travels in the imagined tail. To smoothly connect the two estimations, we adopts the following interpolation:
\begin{equation}
	\Delta'_{\rm rs} = \min\left[\frac{M_{\rm rs}}{4\Gamma_{34}\rho'_{\rm ej}R_{\rm rs}^2}, \Gamma(R_{\rm cd}-R_{\rm rs})\right].
\end{equation}
The density estimation should be correspondingly replaced by
\begin{equation}
    \rho'_{\rm ej}=\frac{M_{\rm ej}}{R_{\rm rs}^2\Delta'_{\rm rs}}.
\end{equation}
This workaround ensures that the mass, density, and volume of the reverse shock region remain self-consistent.

Finally, we stress that the workarounds introduced in this section are relatively unimportant in observation as long as $g \gtrsim 1$. Because we find that after the reverse shock crossing timescale, the emission is primarily dominated by the high-latitude effect.

\subsection{A cold jet with arbitrary angular-radial profile}
To numerically solve the dynamics of the blastwave, we need the boundary conditions $\Gamma_{\rm ej}$ and $n'_{\rm ej}$ at the reverse shock, as demonstrated above. These values must be dynamically derived from the given jet profile during the simulation. This derivation depends on how we define the profile of the jet.

For an axisymmetric jet, the most straightforward way to describe its profile is by specifying its density distribution $\rho'{\rm ej}(r,\theta)$ and velocity distribution $\Gamma{\rm ej}(r,\theta)$. However, this description does not naturally ensure that the tail of the ejecta travels slower than the head; otherwise, internal shocks may develop within the ejecta. Moreover, it does not directly reflect the energy distribution in angular and velocity space, namely $dE/d\Omega$ and $dE/du$. Therefore, we propose an alternative method to describe the jet profile.

In our framework, the jet profile is described by two functions. : $L_{\rm iso}(\tau,\theta)$ and $dE_{\rm iso}(\theta)/d\log u$. The first represents the isotropic equivalent power of an ejecta shell launched by the central engine at time $\tau$ along a given direction. The second describes how the isotropic equivalent energy of the jet is distributed among layers with different speeds. By integrating $dE_{\rm iso}(\theta)/d\log u$ one gets the angular profile of a jet: $E_{\rm iso}(\theta) = 4\pi dE/d\Omega$. These two functions uniquely determine the density and velocity profiles of the jet and naturally ensure that faster layers travel ahead of slower ones. In the rest of this section, we will start from these functions and derive the density and velocity of an ejecta layer when it arrives at the reverse shock. For simplicity, in the discussion below we have omitted the angular profile.

The velocity of an ejecta layer at its launch time $u(\tau)$ is uniquely determined by the jet profile. This can be realized by $L_{\rm iso}(\tau)d\tau=-\frac{dE_{\rm iso}}{d\log u}d\log u$, where the negative sign is because fast layers travel ahead of slow layers. The velocity evolution $u(\tau)$ can then be derived by solving the following ordinary differential equation:
\begin{equation}\label{eq:ejecta_velocity}
    \frac{d\log u}{d\tau}=-L(\tau)\left(\frac{dE_{\rm iso}}{d\log u}\right)^{-1}.
\end{equation}
Because $\frac{d\log u}{d\tau}$ is always negative, the velocity of layers launched at a later time is always slower than that of layers launched at an earlier time, ensuring the absence of internal shocks. Although this equation can be solved analytically in some special cases, we numerically solve it in our approach to adapt general profiles. This process is performed before the blastwave simulation.

To determine the velocity of an ejecta layer when it reaches the reverse shock at simulation time $t$, one must convert between $\tau$ and $t$. This can be achieved by considering the propagation of an ejecta layer:
\begin{equation}\label{eq:time_conversion}
    \tau = t - \frac{r}{\beta c},
\end{equation}
where $\beta$ is the dimensionless velocity of the layer related to $u$. The four-velocity $u$ and launch time $\tau$ of an ejecta layer arriving at the reverse shock at time $t$ and radius $r=R_{\rm rs}$ can thus be numerically solved by combining the above equation with $u(\tau)$.

The density of the ejecta shell can be derived by considering an infinitely thin layer with energy $dE_{\rm iso}$ passing through the reverse shock over an infinitesimal time $dt$, leading to the following result:
\begin{equation}\label{eq:ejecta_density}
    n'_{\rm ej}=\frac{dE_{\rm iso}/dt}{4\pi r^2\beta\Gamma(\Gamma-1)m_p c^3}.
\end{equation}
Here, the power $dE_{\rm iso}/dt$ of the layer is generally different from $L_{\rm iso}(\tau)$ due to velocity gradients that cause the ejecta to stretch during expansion. The conversion relation is $dE_{\rm iso}/dt = L_{\rm iso}(\tau) d\tau/dt$. To find $d\tau/dt$, we differentiate eq. \ref{eq:time_conversion} and get
\begin{align}
    \frac{dt}{d\tau} & = 1 - \frac{r}{\beta\Gamma^2 c}\frac{d\log u}{d\tau} \nonumber \\
    & = 1+\frac{r}{\beta\Gamma^2 c}\left(\frac{dE}{d\log u}\right)^{-1}L(\tau).
\end{align}
Thus, the power of the layer becomes:
\begin{equation}
    \frac{dE_{\rm iso}}{dt}=\left[L(\tau)^{-1}+\frac{r}{\beta\Gamma^2 c}\left(\frac{dE_{\rm iso}}{d\log u}\right)^{-1}\right]^{-1}.
\end{equation}
Inserting this into Eq.\ref{eq:ejecta_density} we can get the ejecta density at the reverse shock. 

\subsection{Reverse shock emission} \label{subsec:emission}
The dynamics of the blastwave are now completely determined by the approach introduced above. The reverse shock emission can be modeled based on a given simulation result.

In this work, we adopt the same synchrotron radiation prescription that we used for the forward shock in our previous study, with the only difference being that the thermodynamic quantities are replaced by those in the reverse shock region. Readers may refer to \citep{2024ApJS..273...17W} for our definition of the synchrotron emissivity $\epsilon'_{\nu'}$ of a fluid element. For the reverse shock region, the intensity of the fluid element in the observer's frame can be calculated as
\begin{equation}
    I_{\nu} = \frac{\epsilon'_{\nu'}}{4\pi}\Delta'_{\rm rs}\delta^3,
\end{equation}
where $\delta$ is the Doppler factor. The observed flux density is given by
\begin{equation}
    F_{\nu} = \frac{(1+z)}{4\pi D_{\rm L}^2} \int  4\pi I_{\nu} R^2 d\Omega,
\end{equation}
where the integral is performed over the equal-arrival-time-surface.

\subsection{Implementation in \texttt{jetsimpy}} \label{subsec:jetsimpy}
The method introduced above to model the evolution of a relativistic jet with a radial profile and the corresponding reverse shock emission has been implemented in \texttt{jetsimpy}. The angular-radial profile of a jet can be defined using tabulated values for $L_{\rm iso}(\tau)$ and $dE_{\rm iso}/d\log u$. This implementation offers unprecedented flexibility in covering a wide range of possible jet profiles, including the simplified profiles widely discussed in the literature, or those derived from GRMHD simulations. Our method involves several numerical processes in addition to the blastwave simulation, such as solving the differential equation in Eq. \ref{eq:ejecta_velocity} and finding the root of Eq. \ref{eq:time_conversion}. We find that their computational costs are subdominant compared to the major simulation, so the overall efficiency is not significantly affected compared to the previous version of the code. The combined flexibility and computational efficiency make the upgraded \texttt{jetsimpy} a valuable tool for the relevant community.

\section{Case studies} \label{sec:cases}
In this section, we apply our model to study several typical cases. First, we revisit the widely discussed reverse shock emission in the early GRB afterglow phase and compare our results with the analytical solutions. In this case the angular profile of the jet is neglected to better investigate the effects of the radial profile and its corresponding reverse shock emission. We then briefly discuss the features of the off-axis reverse shock emission for a structured jet, where both angular and radial profiles are considered. Finally, we demonstrate how our method can be applied to refreshed shock cases, such as kilonova afterglows.

\subsection{Reverse shock emission: the thin and thick shell cases}
Here we apply our model to the early GRB afterglow phase, where reverse shock emission is believed to be responsible for the optical flashes. Ideally, the radial profile can be assumed to be a segment of ejecta with a typical thickness $\Delta_0=\beta_0cT_{90}$, where $\beta_0$ is the initial dimensionless speed of the jet and $T_{90}$ is the typical duration of the GRB. Since the jet may have a radial velocity gradient, such as a radial spreading effect caused by small internal pressure, the jet will expand during its travel. Depending on the extent of this expansion, there are two extreme cases: the thick shell case and the thin shell case.

In the thick shell case, the ejecta thickness is large enough that the radial spreading correction is negligible compared to the initial thickness during the reverse shock crossing phase. In the thin shell case, the radially expanded length dominates the thickness, making the original thickness unimportant. Assuming a power-law external density profile $\rho_0\propto r^{-k}$, one can define the following dimensionless parameter \citep{1995ApJ...455L.143S}:
\begin{equation}
    \xi = \sqrt{\frac{l}{\Delta_0}}\Gamma_0^{-\frac{4-k}{3-k}},
\end{equation}
where $l$ is the Sedov length of the ejecta. When $\xi<1$ the shell is considered to be thick, and when $\xi>1$ the shell is thin. In both cases, there are analytical solutions that can serve as comparisons to our model.

We now apply our model to the two extreme cases, respectively. For simplicity, in this work we consider only a constant ambient density (i.e., $k=0$), although our code is capable of handling the wind-like density profile. The jet profile is assumed to be top-hat, and the observer is assumed to be on-axis.

In the thick shell limit, the velocity gradient in the ejecta shell is negligible, and the shell thickness is dominated by its initial thickness. This means $L_{\rm iso}(\tau)$ is finite and $dE_{\rm iso}/d\log u \sim \infty$. For simplicity, here we assume a constant luminosity profile:
\begin{equation}
    L_{\rm iso}(\tau)=\frac{E_{\rm iso}}{T_{90}}.
\end{equation}
In this limit, eq. \ref{eq:ejecta_density} reduces to
\begin{equation}
    n'_{\rm ej} = \frac{E_{\rm iso}}{4\pi r^2\Delta_0\Gamma_0(\Gamma_0 - 1)m_pc^2}.
\end{equation}
The Lorentz factor of an ejecta layer when it arrives at the reverse shock is simply its initial Lorentz factor $\Gamma_0$.

In the thin shell limit, the shell thickness is dominated by its radially stretched length. A common assumption is that the jet spreads at the local sound speed, and the shell thickness can be estimated as $\Delta\approx r/\Gamma_0^2$. The fact that $\Delta\propto r$ indicates that the shell expansion is homogeneous, which implies a velocity distribution across the different layers of the shell. This distribution could be caused by either natural spreading or an initial velocity gradient. However, we note that most analytic models neglect this velocity gradient and instead assume a uniform velocity for the entire shell. We can estimate the magnitude of this correction in the following way. For a homogeneous shell, the velocity distribution is linear along the radial direction. We can estimate the velocity ratio between the ejecta tail and head by comparing their radii: $\beta_{\rm min}/\beta_0\simeq (r-r/\Gamma_0^2)/r$. This relation gives an estimate of $\beta_{\rm min}\simeq\beta_0^3$, corresponding to a Lorentz factor of $\Gamma_{\rm min}\approx 0.58\Gamma_0$ in the ultra-relativistic limit. Such a variation in Lorentz factor has a significant impact on the blastwave dynamics and the reverse shock emission, implying that this correction is not negligible. In the following, we will discuss the thin shell case with and without this correction. For the description without accounting for the velocity gradient, we refer to it as the ``simplified" thin shell case.

For the "simplified" thin shell case, the situation is similar to that of the thick shell case, except for the expanding shell thickness. The density can be estimated as
\begin{equation}\label{eq:energy_distribution}
    n'_{\rm ej}=\frac{E_{\rm iso}}{4\pi r^2\Delta\Gamma_0(\Gamma_0 - 1)m_pc^2}.
\end{equation}

When the velocity gradient is considered, the ejecta has a finite $dE_{\rm iso}/d\log u$. The jet luminosity can be approximated as $L_{\rm iso}(\tau)\sim \infty$, because the initial thickness of the jet can be regarded negligible. Here, we assume that the energy distribution in the velocity space is
\begin{equation}
    \frac{dE_{\rm iso}}{d\log u}=E_{\rm iso}\left(\frac{\Gamma}{\Gamma_0}\right)^{-2}\frac{\beta}{\beta_0}.
\end{equation}
This profile ensures that the total energy of a shell with thickness $\Delta=r/\Gamma_0^2$ is exactly $E_{\rm iso}$, and that the energy is spatially uniformly distributed. Under this assumption, Eq. \ref{eq:ejecta_density} reduces to 
\begin{equation}
    n'_{\rm ej} = \frac{E_{\rm iso}\beta}{4\pi r^2\Delta\beta_0\Gamma(\Gamma-1)m_p c^2}.
\end{equation}
The velocity of an ejecta layer when it reaches the reverse shock at radius $r$ is simply $\beta=r/ct$, because the expansion is homogeneous. 

\begin{figure}[ht!]
\centering
\includegraphics[width=\columnwidth]{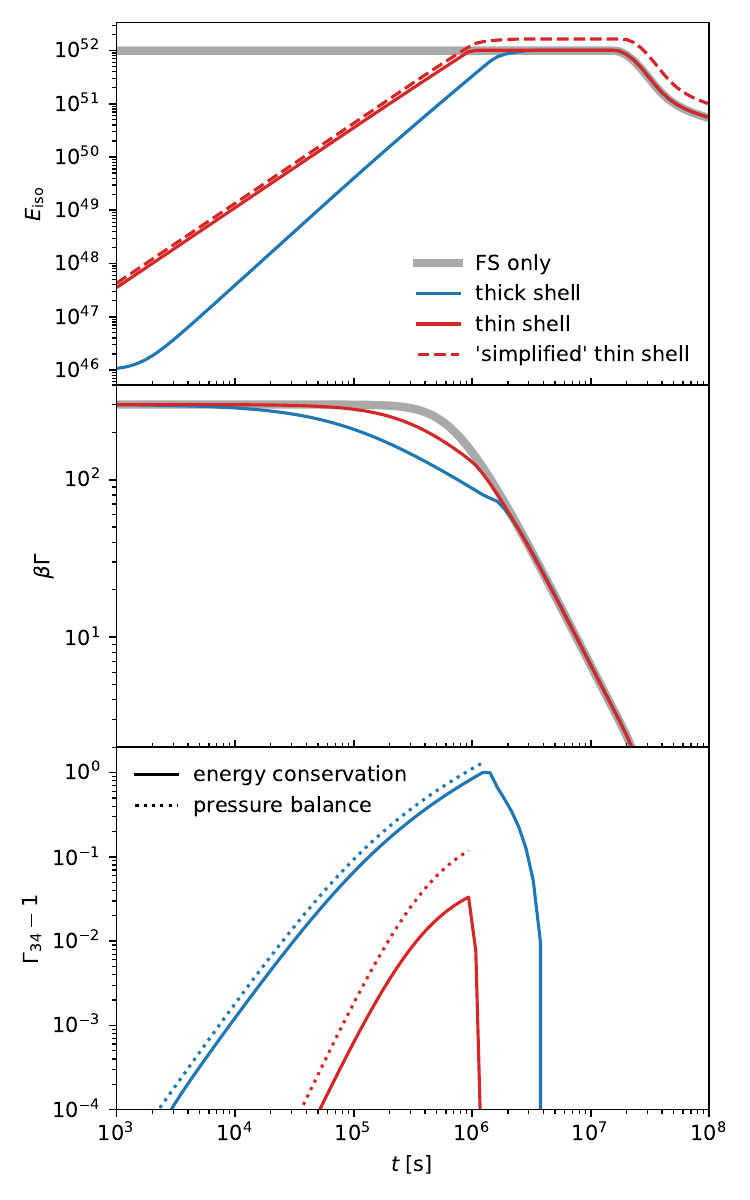}
\caption{The evolution of blastwave properties in the forward-reverse shock system. The parameters for these simulations are: $E_{\rm iso}=10^{52}$ erg, $\Gamma_0=300$, $n_0=1$ cm$^{-3}$, $\theta_c=0.1$ rad, $T_{90}=100$ s, $g=1$. {\it Upper panel}: Evolution of the isotropic equivalent energy for the forward shock-only case, the thick shell case, and the thin shell case. For comparison, the result of the simplified thin shell case is also shown. {\it Middle Panel}: Evolution of the four-velocity. {\it Lower panel}: Evolution of $\Gamma_{34}-1$, representing the energy density in the reverse shock region. For comparison, the prediction based on a pressure balance prescription is also shown.}
\label{fig:thick_and_thin}
\end{figure}

In Fig. \ref{fig:thick_and_thin}, we show the dynamical evolution of the thick and thin shell cases. For comparison, we also show the evolution in the forward shock only case, where all energy have been deposited into the blastwave since the explosion. In the upper panel, the evolution of the energy in the blast region is presented for different cases. The energies for the thin and thick shell cases eventually converge to that of the forward shock-only case, verifying the self-consistency of our approach. In contrast, for the ``simplified" thin shell case, energy conservation is not satisfied, indicating that this description is not self-consistent. Because both the energy and velocity of the ejecta are overestimated, the reverse shock emission predicted in this case as well as in the analytic methods are also overestimated, as will be shown later.

In the middle panel, we show the evolution of the velocity of the blast region. We find that the blastwave starts to decelerate before the crossing time, which is a natural consequence of energy conservation. In particular, we find that the deceleration in the thick shell case is more prominent than in the thin shell case. The initial Lorentz factor of the thick shell drops by a factor of 3 when only 10\% of the energy has been deposited into the blast region. This deceleration affects both $\Gamma$ and $\Gamma_{34}$, leading to more sophisticated light curves in terms of their morphology, as compared to analytic results.

In the lower panel, we further compare our $\Gamma_{34}$ to the one predicted by the pressure balance prescription. In the pressure balance model, $\Gamma_{34}$ is determined by the density ratio between the ejecta and the ambient matter. Assuming a trans-relativistic equation of state, the relation is (e.g., \citealt{2018pgrb.book.....Z})\footnote{In some works (e.g., \citealt{1995ApJ...455L.143S}) the relation is $\frac{n'_{\rm ej}}{n_0}=\frac{(\Gamma-1)(4\Gamma+3)}{(\Gamma_{34}-1)(4\Gamma_{34}+3)}$, which is derived bases on an ultra-relativistic equation of state. However, the two expressions have very little impact on the calculation of $\Gamma_{34}$.}:
\begin{equation}
    \frac{n'_{\rm ej}}{n_0}=\frac{\Gamma^2-1}{\Gamma_{34}^2-1}.
\end{equation}
We calculate $\Gamma_{34}$ using this equation based on the jet dynamics solved in our approach. This ensures that the result is not affected by the deceleration effect. In the lower panel of the figure, we find that the pressure balance prescription overestimates $\Gamma_{34}-1$ compared to the energy conservation prescription. In particular, for the thin shell case, this value is overestimated by a factor of more than 3. As a result, the reverse shock emission is further overestimated.

\begin{figure}[ht!]
\centering
\includegraphics[width=\columnwidth]{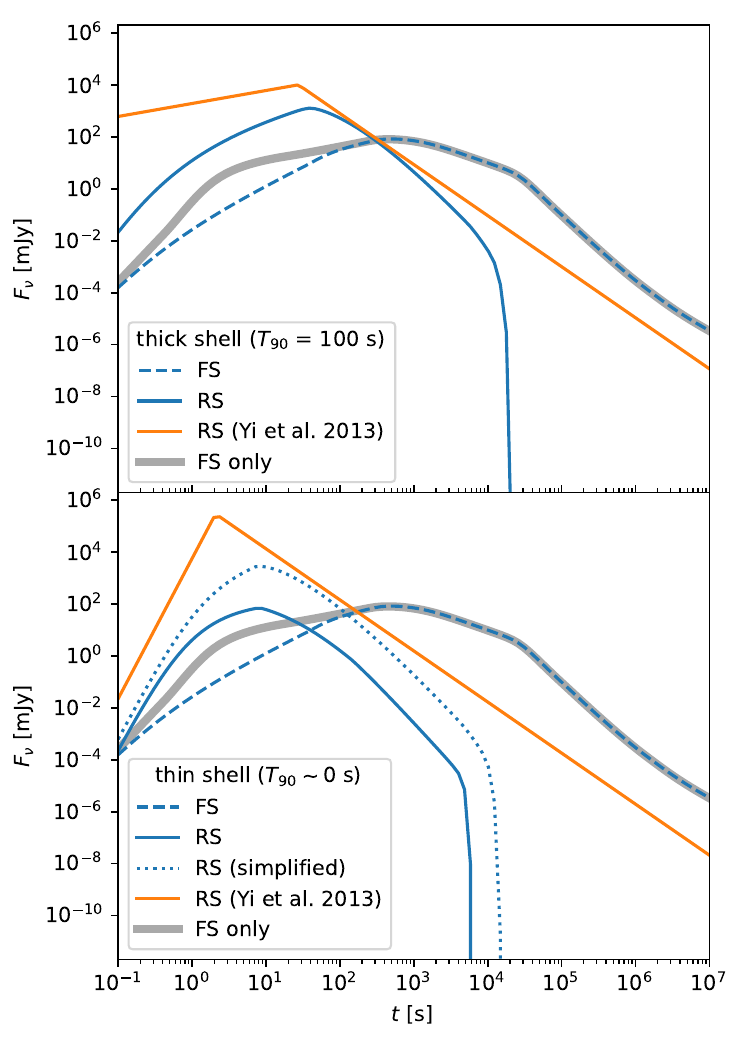}
\caption{The optical light curves of forward and reverse shock emission in the thin and thick shell cases predicted by our model. The simulation parameters are the same as in Fig. \ref{fig:thick_and_thin}. The parameters of the forward shock emission are $\epsilon_{e,\rm fs}=0.1$, $\epsilon_{B,\rm fs}=0.01$, $p_{\rm fs}=2.3$. The parameters of the reverse shock emission are: $\epsilon_{e,\rm rs}=0.1$, $\epsilon_{B,\rm rs}=0.5$, $p_{\rm rs}=2.3$. The luminosity distance is $D_L=400$ Mpc, and the redshift is $z=0.1$. The observing angle is $\theta_v=0$. All light curves are generated at r band ($\lambda=618$ nm) {\it Upper panel}: The thick shell case with $T_{90}=100$ s. For comparison, we also show the analytic light curve based on the model of \citep{2013ApJ...776..120Y}. {\it Lower panel}: The thin shell case. For comparison, we also show the analytic light curve and the ``simplified" thin shell case.}
\label{fig:thick_and_thin_lc}
\end{figure}

In Fig. \ref{fig:thick_and_thin_lc}, we plot the predicted optical light curves of reverse shock emission and show how they differ from the analytic methods. There is a large number of analytical models in the literature, and here we adopt the widely used work of \citet{2013ApJ...776..120Y}. 

For the thick shell case presented in the upper panel, the peak flux of the reverse shock emission is approximately an order of magnitude fainter than the analytic result. This difference arises from the early deceleration of the blastwave and the different predicted values of $\Gamma_{34}$. This level of discrepancy will not significantly affect parameter estimation, as the peak flux is highly sensitive to the initial Lorentz factor.

For the thin shell case shown in the lower panel, the peak flux is more than three orders of magnitude fainter than the analytic prediction. This significant discrepancy arises from the combined effects of the neglected velocity gradient, the overestimated value of $\Gamma_{34}$, and the absence of early deceleration in the pressure balance prescription. Notably, simply neglecting the velocity gradient can lead to an overestimation of the peak flux by two orders of magnitude, as demonstrated in the ``simplified" thin shell case. Such a large difference cannot be reconciled by small adjustments to the parameters. Instead, it indicates that homogeneous shells are unlikely to produce bright reverse shock emission unless additional effects, such as ejecta magnetization, are taken into account. These results help explain why GRB afterglows with reverse shock emission detected are rare.

In both cases, the light curves of the reverse shock emission exhibit a curved morphology, in contrast to the power-law behavior predicted by the analytic model. The forward shock emission is also reduced during the coasting phase. These effects are caused by the deceleration of blastwaves during the reverse shock crossing period, which reduces the luminosity during the later stage of reverse shock crossing.

\subsection{Off-axis reverse shock emission from a structured jet}
In this section, we briefly discuss the off-axis reverse shock emission for a structured jet as predicted by our model. For an off-axis observer, the afterglow emission can differ significantly from that of an on-axis observer. Both the forward and reverse shock emission from the jet core are initially beamed away from the line of sight and gradually become visible as the jet decelerates. For the forward shock emission, this process leads to a light curve slope that depends on the observing angle. The situation for the reverse shock emission is more complex. The jet material in the high-latitude regions may have a small Lorentz factor, corresponding to the thin shell case, while the jet core may have a large Lorentz factor, corresponding to the thick shell case. For an off-axis observer, it is possible to observe a transition from thin shell to thick shell cases \citep{2025arXiv250212757A}.

Here, we consider a power-law jet structure, as suggested by previous studies (e.g., \citealt{2003ApJ...586..356Z,2011ApJ...740..100B,2021MNRAS.500.3511G}). The energy and Lorentz factor are described by the following profiles:
\begin{align}
    & E_{\rm iso}(\theta) = E_c\left[1+(\theta/\theta_c)^2\right]^{-s/2}, \\
    & \Gamma_0(\theta) = (\Gamma_c-1)\left[1+(\theta/\theta_c)^2\right]^{-s/2} + 1,
\end{align}
where the half-opening angle is $\theta_c=0.1$. In this study, the slope is assumed to be $s=4$. This value determines the angle at which the ejecta shell transitions between the thin and thick shell cases, and it does not affect our conclusion. For our selected values, the ejecta shell is thick for $\theta<0.21$ and thin for $\theta>0.21$.

To smoothly connect the thin and thick shell cases, we propose a ``unified" jet radial profile. We assume that the jet temporal luminosity follows the thick shell case (i.e., $L_{\rm iso}(\tau)=E_{\rm iso}/T_{90}$), while the jet energy distribution in velocity space follows the thin shell case (see eq. \ref{eq:energy_distribution}). For this specific jet profile, the temporal velocity evolution at launch time can be solved analytically:
\begin{equation}\label{eq:unified_velocity}
    \beta=\beta_0(1-\frac{1}{\Gamma_0^2}\frac{\tau}{T_{90}}).
\end{equation}
The density of an ejecta layer when it reaches the reverse shock is computed using eq. \ref{eq:ejecta_density}, and its velocity should be numerically solved using Eq. \ref{eq:time_conversion} and eq. \ref{eq:unified_velocity}.

\begin{figure}[ht!]
\centering
\includegraphics[width=\columnwidth]{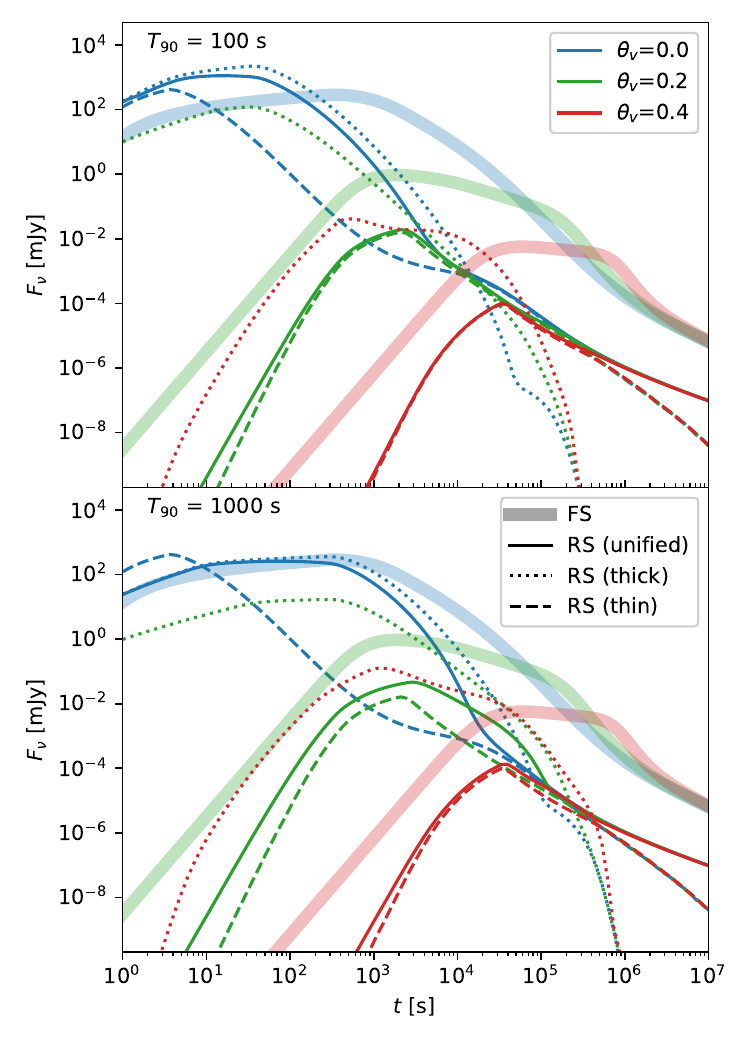}
\caption{Off-axis reverse shock emission at optical waveband for a structured jet. The angular structure follows a power-law profile with a half-opening angle of $\theta_c=0.1$ and a slope of $s=4$.} The parameters for the jet simulation are: $E_c=10^{52}$ erg, $\Gamma_c=300$, $n_0=10$ cm$^{-3}$, and $g=1$. The duration of the jet is $T_{90}=100$ s in the upper panel and $T_{90}=1000$ s in the lower panel. The parameters and frequencies for the forward and reverse shock emissions are the same as in Fig. \ref{fig:thick_and_thin_lc}.
\label{fig:misaligned}
\end{figure}

In Fig. \ref{fig:misaligned}, we show the synthetic light curves for both forward and reverse shock emissions of a structured jet observed from various directions. For comparison, we also present the reverse shock emission predicted by the pure thin and thick shell models. The parameters are listed in the figure caption. 

In the upper panel, the duration of the jet is $T_{90}=100$ s. We find that the on-axis afterglow mostly follows a thick shell case, while the off-axis afterglow primarily follows a thin shell case. In the lower panel, the duration of the jet is increased to $T_{90}=1000$ s. When $\theta_v=0.2$, a thin-to-thick transition can be observed. However, its morphology cannot be well distinguished from that of a pure thin shell case.

This transition is difficult to observe because the peak time of the afterglow emission for a misaligned observer is typically later than the reverse shock crossing time of the thick shell regions, unless the jet duration is extremely long. The observing angle must also be carefully adjusted. It should not be too small, where the thick shell emission dominates, nor too large, where the peak time occurs after the thick shell crossing time. Therefore, we conclude that the conditions required to find a thin-to-thick transition are quite extreme, posing a challenge to the observation.

\subsection{Long-lived reverse shock and refreshed shocks: Kilonova afterglow} \label{subsec:KN afterglow}
In addition to the reverse shock emission, an extended jet radial profile can also play a role in energy injection. In certain situations, a long lasting energy input can significantly increase the energy of the blastwave, even if the corresponding reverse shock is very weak. In these cases, the impact of the radial profile is primarily reflected in the emission from the forward shock. This phenomenon is generally referred to as a refreshed shock. Recently, this model has been applied to kilonova afterglows, where the ejecta energy is significantly extended in the velocity space. It can also explain the early plateau phase or the late-time rebrightening observed in some GRB afterglows. As we will show below, the refreshed shock model can be naturally realized within our framework.

In this work, we follow the common assumption used in modeling kilonova afterglows, where the ejecta expansion is assumed to be homogeneous and isotropic. The energy distribution for ejecta layers with different four-velocities is assumed to be
\begin{equation}
    \frac{dE_{\rm iso}}{d\log u}=\alpha E_{\rm iso}\left(\frac{u}{u_{\rm min}}\right)^{-\alpha},
\end{equation}
where $u_{\rm min}$ is the minimum four-velocity of the ejecta. This assumption is mathematically equivalent to the form commonly used in the literature: $E(>u_{\rm min})\propto u^{-\alpha}$. Because the shell is homogeneous, this scenario is similar to the thin shell case, except that the velocity distribution can be very broad: $u_{\rm min}\ll u_{\rm max}$. The initial luminosity can be approximated as $L_{\rm iso}(\tau)\sim \infty$. 

For this specific type of radial profile, previous studies have found power-law approximations to the predicted light curves $F_{\nu}\propto t^{s}$, where the slope is \citep{2015MNRAS.448..417B,2018MNRAS.473L.121K}
\begin{equation}\label{eq:KN_afterglow_slope}
	s = 
	\begin{cases}
		\frac{6(p-1)-3\alpha}{8+\alpha} & {\rm relativistic} \\
		\frac{15p-21-6\alpha}{10+2\alpha} & {\rm Newtonian} \\
	\end{cases}.
\end{equation}

Unlike the early afterglow phase, in this case the reverse shock emission is expected to be relevant only in the radio waveband. This is because during the early afterglow phase, the blastwave is ultra-relativistic, and the reverse shock emission can be boosted to the optical waveband. However, the kilonova ejecta are only mildly relativistic or Newtonian, and the reverse shock emission is most likely detected in the radio waveband.

\begin{figure}[ht!]
\centering
\includegraphics[width=\columnwidth]{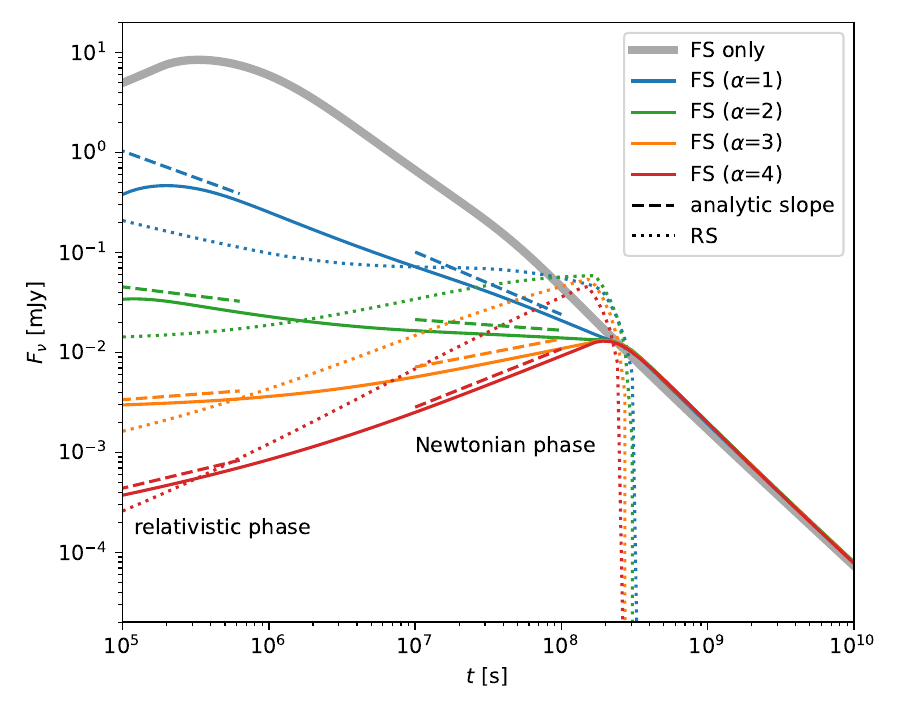}
\caption{The kilonova afterglow emission predicted by our model. For comparison, the forward shock only case is also shown to demonstrate the consistency of our model. The parameters for the ejecta are $E_{\rm iso}=10^{51}$ erg and $u_{\rm min}=0.2$. The parameters for the forward and reverse shock emissions are the same as in Fig. \ref{fig:thick_and_thin_lc}. All light curves are generated at frequency $\nu=3$ GHz.}
\label{fig:cocoon_afterglow_lc}
\end{figure}

In Fig. \ref{fig:cocoon_afterglow_lc}, we present the kilonova afterglow light curve at the radio waveband predicted by our model. The parameters are listed in the figure caption. The analytical slope based on this result is also plotted for comparison. To further verify the self-consistency of our calculation, we include the forward shock only case, where all of the energy is deposited into the blastwave at the beginning. The convergence of all light curves verifies the self-consistency of our approach. Our calculations show a good agreement with the analytical results, indicating that the inclusion of reverse shock doesn't significantly change the blastwave dynamics compared to the works that only consider forward shocks (e.g., \citealt{2019MNRAS.487.3914K}). This is likely because the deceleration due to reverse shock is less significant if the expansion of the ejecta is homogeneous, as we have discussed in the thin shell case (see Fig. \ref{fig:thick_and_thin}). 
We also find that the luminosity of the reverse shock emission at radio wavebands could exceed that of the forward shock emission for certain parameters (e.g., $\epsilon_B=0.5$ adopted in this work).

It is worth noting that in the Newtonian phase of the kilonova afterglow, corrections to the synchrotron radiation prescription are required when the minimum electron Lorentz factor approaches $\gamma_m \sim 1$, a regime commonly referred to as the ``deep Newtonian phase" \citep{2003MNRAS.341..263H,2013ApJ...778..107S}. In this phase, the slope could be different from that in Eq. \ref{eq:KN_afterglow_slope} (e.g., \citealt{2025A&A...693A.108K}). This effect is not taken into account in the current modeling. We leave it for future works.

\section{Application to GRB 990123 afterglow} \label{sec:grb990123}
To test our modeling of the jet radial profile against real-world data, especially its predicted reverse shock emission, we apply our model to fit the afterglow data of GRB 990123. This GRB exhibited a very bright early optical flash, which requires extreme parameters in analytical reverse shock models (see also \citet{2002ChJAA...2..449F} and \citet{2003ApJ...595..950Z}). As shown in previous sections, the reverse shock emission predicted by our model is generally fainter than that of the analytic methods. Therefore, this GRB can serve as a test for our model under extreme conditions.

In this fitting, the optical data are collected from \citet{1999Natur.398..400A} and \citet{1999Sci...283.2069C}. These data have been converted to the AB magnitude system based on \citet{2018ApJS..236...47W}\footnote{\url{https://mips.as.arizona.edu/~cnaw/sun.html}}, and they are corrected for Galactic extinction \citep{2011ApJ...737..103S}
\footnote{\url{https://irsa.ipac.caltech.edu/applications/DUST/}}. The X-ray data set is taken from \citet{2005A&A...438..821M}. The radio data set for this GRB is very limited, consisting of only a single significant detection. This point is excluded from the fitting, because radio emission may be influenced by synchrotron self-absorption and interstellar scintillation, which may introduce bias into the results.

We assume a thick shell model for the radial profile of the jet, as suggested by pervious studies \citep{2002ChJAA...2..449F,2000MNRAS.319.1159W}. For a thick shell model, the parameter set to determine the dynamical evolution of the jet includes the isotropic equivalent energy $E_{\rm iso}$, the ambient medium number density $n_0$, the initial Lorentz factor of the jet $\Gamma_0$, the half opening angle of the jet $\theta_c$, the engine activity timescale $T_{90}$, and the reverse shock cooling coefficient $g$. The forward shock emission depends on the following microphysical parameters: the electron energy fraction $\epsilon_{e,\rm fs}$, the magnetic field energy fraction $\epsilon_{B,\rm fs}$, and electron power index $p_{\rm fs}$. Similarly, the microphysical parameters in the reverse shock region are $\epsilon_{e,\rm rs}$, $\epsilon_{B,\rm rs}$, and $p_{\rm rs}$.
Finally, the cosmological parameters of \cite{2014ApJ...794..135B} are adopted to calculate the luminosity distance of GRB 990123 at $z=1.6$.

\begin{table*}[t]
	\centering
	\caption{GRB 990123 afterglow fitting parameters.}
        \label{tab:param}
	\begin{tabular}{cccc}
		\hline
		Parameters & Descriptions & Prior distributions & 1-$\sigma$ fitting uncertainty \\
		\hline
		$\log_{10}(E_{\rm iso}/{\rm erg})$ & Jet isotropic equivalent energy & 55 & - \\
            $\log_{10}(n_0/{\rm cm^{-3}})$ & Circumburst medium number density & [-3, 1] & $-1.726^{+0.108}_{-0.068}$ \\
            $\log_{10}\Gamma_0$ & Jet initial Lorentz factor & [1, 4] & $2.742^{+0.008}_{-0.013}$ \\
            $\theta_{\rm c}$ [rad] & Jet half opening angle & [0, $\pi/2$] & $0.045^{+0.002}_{-0.001}$ \\
            $\theta_{\rm v}$ [rad] & Observing angle & 0.0 & - \\
            $\log_{10}\epsilon_{e, \rm fs}$ & Forward shock electron energy fraction & [-6, 0] & $-1.584^{+0.044}_{-0.028}$ \\
            $\log_{10}\epsilon_{B, \rm fs}$ & Forward shock magnetic field energy fraction & [-6, 0] & $-3.921^{+0.061}_{-0.091}$ \\
            $p_{\rm fs}$ & Forward shock electron power index & [2, 3] & $2.510^{+0.024}_{-0.017}$ \\
            $\epsilon_{e, \rm rs}$ & Reverse shock electron energy fraction & 0.5 & - \\
            $\log_{10}\epsilon_{B, \rm rs}$ & Reverse shock magnetic field energy fraction & [-6, 0] & $-0.063^{+0.042}_{-0.074}$ \\
            $p_{\rm rs}$ & Reverse shock electron power index & 2.5 & - \\
            $g$ & Cooling coefficient of the reverse shock region & [0.1, 2] & $1.934^{+0.044}_{-0.086}$ \\
            $T_{90}$ [s] & Engine activity timescale & [2, 70] & $13.819^{+0.075}_{-0.074}$ \\
            \hline
	\end{tabular}
\end{table*}

It is easy to realize that the limited observational data set is insufficient to tightly constrain all the parameters mentioned above, and strong parameter degeneracies are expected in the fitting process. Therefore, we reasonably fix some parameters to improve fitting efficiency without affecting the validity of the results. Because the early optical flash is observed in only one waveband, it is unlikely that the spectral power-law index and break frequencies of the reverse shock can be well constrained. We thus assume $p_{\rm rs}=2.5$ and $\epsilon_{e,\rm rs}=0.5$, leaving only $\epsilon_{B,\rm rs}$ as a free parameter. For the forward shock, the spectral break associated with $\nu_m$ is not observed in the optical band, which will lead to a degeneracy among $n_0$, $\epsilon_{e,\rm fs}$, and $\epsilon_{B,\rm fs}$. The uncertainties in these parameters further propagate to $E_{\rm iso}$ due to the degeneracy between $n_0$ and $E_{\rm iso}$. To break these degeneracies, we fix $E_{\rm iso}=10^{55}$ erg. This value is motivated by the isotropic-equivalent energy of the prompt emission, $E_{\rm \gamma,iso}\approx 3\times 10^{54}$ erg, and the expectation that GRBs generally have high radiation efficiency. Finally, we fix $\theta_v=0$ because the event is most likely observed on-axis. With these assumptions, the dimension of the parameter space is reduced to just nine. We find that fixing these parameters can mostly break the degeneracies in the fitting. The fitting parameters are summarized in Table \ref{tab:param}.

\begin{figure}[ht!]
\centering
\includegraphics[width=\columnwidth]{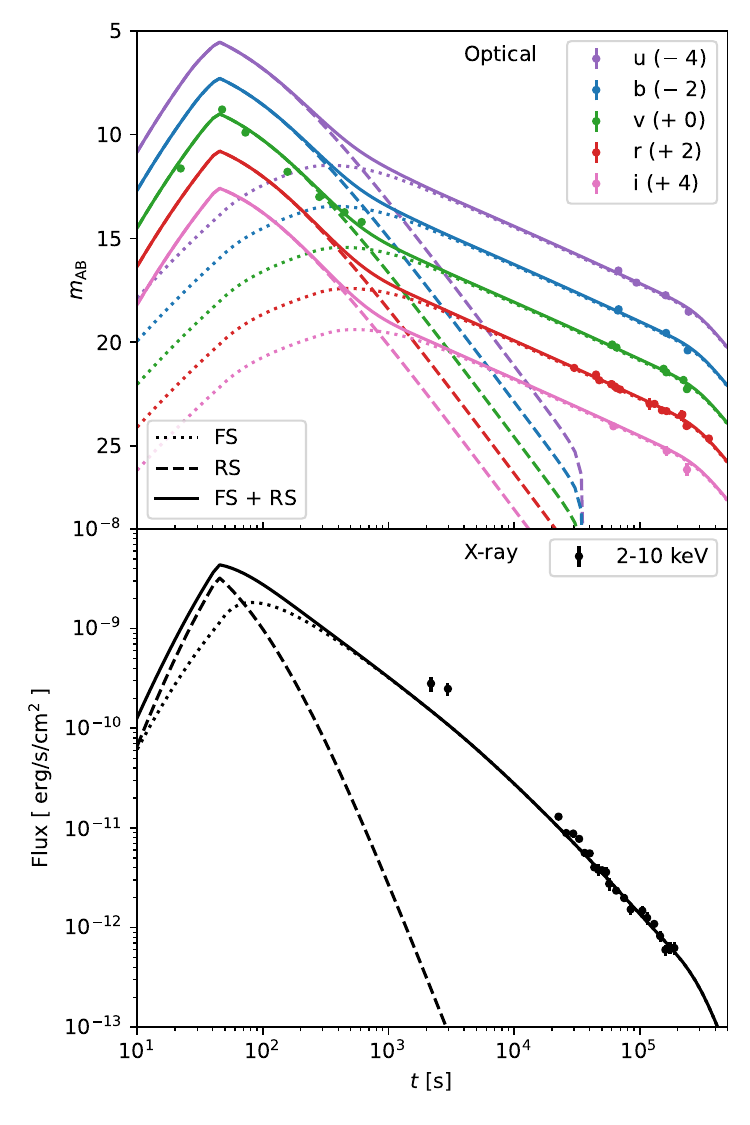}
\caption{The best fitting optical and X-ray light curves for GRB 990123 afterglow. The optical data are converted to AB magnitude and are corrected for galactic extinction.}
\label{fig:best_fit_lc}
\end{figure}

\begin{figure*}[t]
\centering
\includegraphics[width=\textwidth]{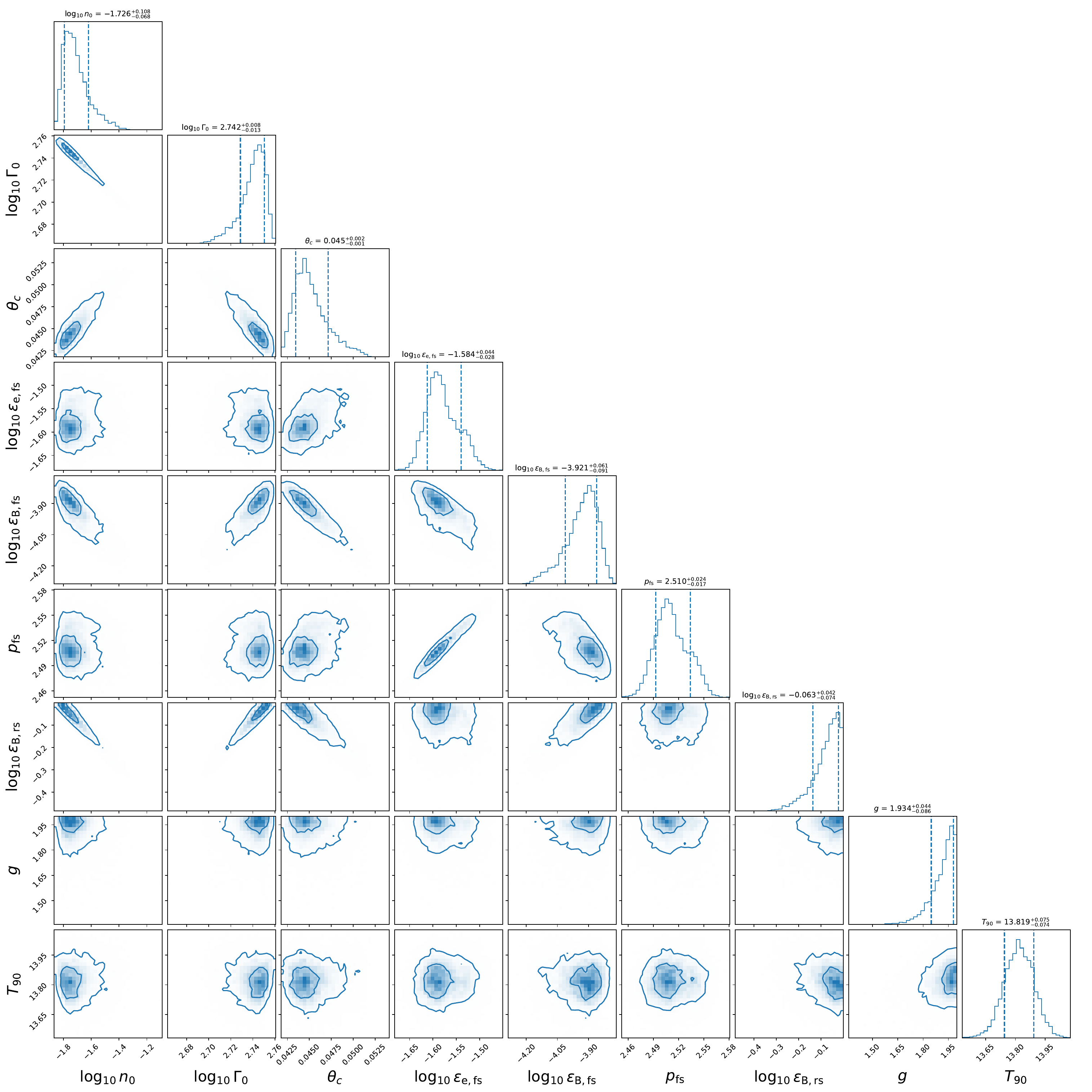}
\caption{The parameter estimation results for GRB 990123 afterglow. The parameter uncertainties are in 1-$\sigma$ level, and the contour plots show 1-$\sigma$ and 2-$\sigma$ levels.}
\label{fig:corner}
\end{figure*}

In this fitting, we perform a Bayesian parameter estimation. The likelihood function is defined as $\mathcal{L}=\exp[-(\chi_{\rm opt}+\chi_x)/2]$, where $\chi_{\rm opt}$ and $\chi_x$ are the reduced chi-square values for the optical and X-ray data sets, respectively. The prior distributions for all parameters are assumed to be uniform, and their ranges are listed in Table \ref{tab:param}. The sampling of Posterior probability distribution is performed by \texttt{bilby} \citep{bilby_paper}, where we have used the embedded \texttt{pymultinest} sampler \citep{2014A&A...564A.125B}. The sampling results are presented in Figure \ref{fig:corner}, and the parameter uncertainties are summarized in Table \ref{tab:param}. The best-fitting light curves are shown in Figure \ref{fig:best_fit_lc}.

Our fitting shows good agreement with the observational data, as illustrated in Fig. \ref{fig:best_fit_lc}. The minor discrepancy observed in the first two X-ray data points is probably due to the inaccuracy in estimating the cooling frequency $\nu_c$ under the global cooling approximation, because the fitting shows that the x-ray waveband is above $\nu_c$. A more accurate estimation of the synchrotron cooling frequency would require numerical solving of the Fokker-Planck equation. The estimated initial Lorentz factor of the jet is approximately $\Gamma_0\sim 500$, which is consistent with previous studies. The inferred magnetic energy fraction in the reverse shock region is $\epsilon_B\sim 1$, suggesting that the shocked ejecta is at least mildly magnetized (see also \citet{2002ChJAA...2..449F} and \citet{2003ApJ...595..950Z}). The inferred $\epsilon_B$ violates the limit $\epsilon_e+\epsilon_B\lesssim 1$, but the violation is not severe. This problem can be alleviated if the magnetization of the ejecta is considered for more accurate modeling, which we leave to future work. The estimated $T_{90}$ differs from the duration inferred from the prompt emission, $T_{90}\sim$ 60 s. However, the late-time emission in the prompt phase may be extended emission, possibly originating from inverse Compton scattering in the forward shock region. In fact, the duration of the main emission pulses is approximately 20 s (see Fig. 2 in \citet{1999Natur.398..400A}), which is consistent with our result. The estimated $g$ shows strong evidence for $g>1$, indicating that the shocked ejecta cools rapidly after the reverse shock crossing.

This fitting has validated our approach, indicating that our model is capable of explaining the early optical flashes observed in some of the GRB afterglows, even if the flashes are very bright. We stress that this fitting is not a comprehensive study of the GRB 990123 afterglow, but only a test to our code. 

\section{Summary and Discussion} \label{sec:summary}
In this work, we have further developed our previous code, \texttt{jetsimpy}, which models the GRB afterglow produced by structured jets. In addition to the previously studied angular profile, we introduce a stratified radial profile for the jet. This allows the code to model jets with arbitrary axisymmetric two-dimensional configurations. The presence of a stratified radial profile leads to the formation of a reverse shock and hydrodynamic energy injection. In studying the dynamics of the forward–reverse shock system, we assume energy conservation of the blast wave instead of pressure balance at the contact discontinuity. Under this framework, we study the features of on-axis and off-axis reverse shock emission predicted by our model. We have also shown that our model can be naturally applied to kilonova afterglows. Our major findings are summarized as follows:

\begin{enumerate}
  \item Within the framework of energy conservation, we find that the blastwave decelerates earlier than the reverse shock crossing timescale, especially in the thick shell case. This effect leads to a reduced luminosity before the crossing timescale. The forward shock emission is also suppressed during the coasting phase.
  
  \item We find that the energy density of the reverse shock region, characterized by $\Gamma_{34}-1$, has been overestimated under the pressure balance framework, especially in the thin shell case.
  
  \item In the thin shell case, analytical models have neglected the velocity gradient throughout the ejecta. We find that this effect is non-negligible, as it results in self-inconsistent reverse shock modeling.
  
  \item Combining all the effects mentioned above, we find that the peak luminosity of the reverse shock emission in thin shell cases has been significantly overestimated by analytical models, by more than three orders of magnitude. In thick shell cases, the overestimation is approximately one order of magnitude. These findings help explain why GRB afterglow events with a significant reverse shock component detected are rare.

  \item For structured jets, the reverse shock emission mostly follows either a thick or thin shell case, depending on the energy at the initial visible angle. For off-axis observers, it is possible to observe a thin-to-thick transition; however, we find that this transition is very difficult to detect in practice, as it requires a very long jet duration and a fine-tuned observing angle.

  \item In modeling the kilonova afterglow, we find that the predicted slopes of the light curves are consistent with analytical results. We also find that in radio wavebands the reverse shock emission could exceed that of the forward shock emission if the microphysical parameters (i.e., $\epsilon_e$ and $\epsilon_B$) in the reverse shock region are reasonably high.

  \item In fitting the GRB 990123 multi-waveband afterglow, we find that the ejecta are likely at least mildly magnetized. A reverse shock model considering ejecta magnetization may better explain the optical flash observed in this event.
\end{enumerate}

In addition to the cases discussed in this work, our model can also be applied to scenarios involving energy injection, such as the early plateau phase and late-time rebrightening observed in some GRB afterglows. The ambient density profile can also be arbitrary, such as a stellar wind profile following $n_0\propto r^{-2}$. These cases will be explored in future works.

In the current stage, magnetization is not included in our model. However, since some GRB jets are thought to be Poynting-flux dominated, this effect may play an important role in modeling the reverse shock. Incorporating ejecta magnetization may increase or suppress the brightness of the emission (e.g., \citealt{2004A&A...424..477F,2005ApJ...628..315Z,2008A&A...478..747G}). We propose to further develop our code and include this feature in a future work.

\section*{acknowledgments}
The authors thank Jia Ren, Yun Wang, Paz Beniamini, and Dimitrios Giannios for their insightful discussions and comments on reverse shock emission. We also appreciate the anonymous referee for the valuable suggestions. This work is supported by the Natural Science Foundation of China (NSFC) under Grant Nos. 12321003, 12225305, 12473049, the Strategic Priority Research Program of the Chinese Academy of Sciences, Grant No. XDB0550400, and the National Key R\&D Program of China (2024YFA1611704).

%

\vspace{5mm}


\software{Jetsimpy\citep{2024ApJS..273...17W},  
          Matplotlib \citep{Hunter:2007}, Bilby\citep{bilby_paper}, pymultinest\citep{2014A&A...564A.125B}
          }



\appendix

\section{Eigenvalues of the Partial Differential Equations}\label{appendix:numerical_scheme}
To construct the numerical scheme for solving the evolution equation (eq. \ref{eq:evolution_equation}), one needs the eigenvalues of the Jacobian matrix $\partial \bf{F}/\partial\bf{U}$ which can be found in Appendix A of our previous work \citep{2024ApJS..273...17W}. The mathematical form of the Jacobian matrix and its eigenvalues remain the same if they are expressed in terms of $\partial P_b/\partial E_b$, $\partial P_b/\partial M_{\rm fs}$, and $\partial P_b/\partial M_{\rm rs}$. However, the inclusion of reverse shock thermal energy in $E_b$ leads to modifications in the calculation of these variables. To derive these variables, we first differentiate $P_b=\beta^2M_{\rm fs}/3$ and derive $dP_b$ in terms of $d\beta$ and $dM_{\rm fs}$. Note that $\beta$ is an implicit function of $E_b$, $M_{\rm fs}$, and $M_{\rm rs}$, as defined by Eq. \ref{eq:energy}. We then differentiate this equation and derive $d\beta$ in terms of $dE_b$, $dM_{\rm fs}$, and $dM_{\rm rs}$. Replacing the $d\beta$ term in $dP_b$ by this result, we can derive the required variables. After some algebra, we obtain the following results:
\begin{align}
    & \frac{\partial P_b}{\partial E_b} = \frac{B}{A}, \\
    & \frac{\partial P_b}{\partial M_{\rm fs}} =  \frac{1}{3}\beta^2 - \Gamma^2(1+\frac{1}{3}\beta^4)\frac{B}{A}, \\
    & \frac{\partial P_b}{\partial M_{\rm rs}} = -\Gamma\Gamma_{34}(1+\frac{1}{3}\beta^2\beta_{34}^2)\frac{B}{A},
\end{align}
where $A$ and $B$ are defined by
\begin{align}
    & A = \frac{2}{3}\beta(4\Gamma^4-1)M_{\rm fs}+\beta\Gamma^2\left[\Gamma\Gamma_{34}\left(1+\frac{1}{3}\beta^2\beta_{34}^2\right)\left(1-\frac{\beta_{34}}{\beta}\right)+\frac{2}{3}\beta\beta_{34}\left(\frac{\beta_{34}\Gamma_{34}}{\beta\Gamma}-\frac{\Gamma}{\Gamma_{34}}\right)\right]M_{\rm rs}, \\
    & B = \frac{2}{3}\beta M_{\rm fs}.
\end{align}
These expressions are reduced to those in \citet{2024ApJS..273...17W} in the absence of reverse shock, that is, when $\Gamma_{34}=1$ and $\beta_{34}=0$.



\bibliography{sample631}{}
\bibliographystyle{aasjournal}


\end{CJK*}
\end{document}